\documentclass[%
reprint,
superscriptaddress,
amsmath,
amssymb,
aps,
prl,
twocolumn,
floatfix,
]{revtex4-2}
\usepackage{physics,hyperref,booktabs,bm,mathrsfs}
\usepackage[linesnumbered,ruled,vlined]{algorithm2e}
\hypersetup{colorlinks=true,breaklinks,urlcolor=blue,linkcolor=blue,citecolor=blue}

\usepackage[labelformat=simple]{subcaption}
\usepackage{ragged2e}
\DeclareCaptionJustification{justified}{\justifying}
\usepackage{graphicx}
\newcommand{\inputtikz}[1]{\includegraphics{#10.pdf}}

\makeatletter
\def\@listcomma@comma{}
\makeatother

\begin{document}
\title{Solving Sharp Bounded-error Quantum Polynomial Time Problem by Evolution methods}
\author{Zhen Guo}
\email[Email:]{guozhen1201@gmail.com}
\affiliation{State Key Laboratory of Low Dimensional Quantum Physics, Department of Physics, Tsinghua University, Beijing 100084, China}
\author{Li You}
\affiliation{State Key Laboratory of Low Dimensional Quantum Physics, Department of Physics, Tsinghua University, Beijing 100084, China}
\affiliation{Frontier Science Center for Quantum Information, Beijing, China}
\affiliation{Hefei National Laboratory, Hefei, 230088, China}

\date{\today}
\begin{abstract}
    Counting ground state degeneracy of a $k$-local Hamiltonian is important in many fields of physics.
    Its complexity  belongs to the problem of sharp bounded-error quantum polynomial time (\#BQP) class and few methods are known for its solution.
    Finding ground states of a $k$-local Hamiltonian, on the other hand, is an easier problem of Quantum Merlin Arthur (QMA) class, for which many efficient methods exist.
    In this work, we propose an algorithm of mapping a \#BQP problem into one of finding a special ground state of a $k$-local Hamiltonian.
    We prove that all traditional methods, which solve the QMA problem by evolution under a function of a Hamiltonian, can be used to find the special ground state from a well-designed initial state, thus can solve the \#BQP problem.
    We combine our algorithm with power method, Lanczos method, and quantum imaginary time evolution method for different systems to illustrate the detection of phase boundaries, competition between frustration and quantum fluctuation, and potential implementations with quantum circuits.
\end{abstract}

\maketitle

\paragraph*{Introduction.}

Solving for the ground states of a $k$-local Hamiltonian is an essential problem in quantum mechanics.
It belongs to the complexity class of quantum Merlin Arthur (QMA)~\cite{Aharonov2002}.
Many methods are designed for either classical~\cite{Schollwoeck2011,Verstraete2004b,Daley2004,Nishio2004,Vidal2007,Foulkes2001,Austin2012,Zwolak2004} or quantum~\cite{Farhi2014,Peruzzo2014,McArdle2019,Motta2019} computers.
Some of them essentially find the ground state in the form of a function of Hamiltonian acting on an initial state, e.g.\ imaginary time evolution~\cite{McArdle2019,Motta2019,Zwolak2004}, power methods~\cite{Mises1929}, and Lanczos method~\cite{Lanczos1950,Komzsik2003}, which will be called evolution-based methods.
In theory, these methods are immune to being trapped at local minima.
In contrast, counting the ground state degeneracy belongs to the class of sharp bounded-error quantum polynomial time (\#BQP)~\cite{Brown2011}, which belongs to a harder complexity and few methods are known for its solutions.

The degeneracy of ground state is an important property to many fields in physics.
The ground state may possess less symmetry than its Hamiltonian.
Such spontaneous symmetry breaking causes the ground state to be degenerate, which plays a significant role in research on phase transition~\cite{Beekman2019}.
Topological phases, attracting increased interests~\cite{Yu2020, Greiter2014, Hung2015, Jiang2020, Kennedy1990, Kitaev2001, Kitaev2009, Klassen2015, Levin2006, Ling2022, Mahyaeh2020, Misguich2002, Rao2014, Sarma2015, Sreejith2011, Sun2016, Wei2022, Schmitz2013, Pollmann2010, Mahyaeh2018} in recent years, are differentiated by the existence of topological zero modes instead of different symmetries and also result in degenerate ground states.
Another related topic concerns systems with large ground state degeneracies at zero temperature due to frustration. Related discussions date back to the early days when the physical meaning of residual entropy was debated~\cite{Pauling1935}, well-known examples include spin glass phase and spin-liquid phase~\cite{Goldstein2008, Gupta2008, Jaeckle1984, Jurcisinova2020, Kirkpatrick1977, Mezard1984, Cuevas2016, Zhou2017}.

In this work, we show that the problem of counting the ground state degeneracy (a \#BQP problem) maps into one of solving for a special ground state, and we prove that all evolution-based methods can be employed to find such special ground state from a well-designed initial state, and hence can solve the original \#BQP problem.
Our algorithm is combined with the matrix product states (MPS) for 1D systems, Lanczos method for 2D frustrated systems, and quantum imaginary time evolution (QITE) for 3D ab-initio Hamiltonians to illustrate its potential applications to research on phase detection, frustration, and quantum computation.

\paragraph*{Mapping \#BQP into special QMA.}
Given an orthonormal complete operator basis under the Hilbert-Schmidt inner product $X\cdot Y\triangleq\Tr (X^\dag Y)$, e.g.~the Pauli basis $O_\alpha^{pq}$ with $p, q$ the index of matrix elements and $O_0=\sigma_0/\sqrt2, O_1=\sigma_x/\sqrt2, O_2=\sigma_y/\sqrt2, $ and $O_3=\sigma_z/\sqrt2$, an arbitrary Hamiltonian acting on $N$ qubits can be expanded into the form
\begin{equation}
        H =\sum\limits_{\{\alpha\}} c_{\{\alpha\}} \bigotimes_{k=1}^{N} O_{\alpha_k},
\end{equation} with $\alpha_k\in\{0,1,2,3\}$.
We can define a corresponding super Hamiltonian $\tilde{H}$ acting on $N$ ququarts (tilde indicates the quantity for ququarts).
\begin{equation}
    \tilde{H} \triangleq\sum\limits_{\{\alpha\}} c_{\{\alpha\}} \bigotimes_{k=1}^{N} \tilde{O}_{\alpha_k},\label{eq:tildeH}
\end{equation} with matrix elements of $\tilde{O}_\alpha$ defined as\begin{equation}\label{equ:basetrans}
    {\tilde{O}_\alpha^{\beta\gamma}} \triangleq \Tr (O_\beta^\dag O_\alpha O_\gamma).
\end{equation}
The coefficients of the super Hamiltonian $\tilde{H}$ are identically the same as the original one except its basis $O$ is substituted by one acting on ququarts $\tilde{O}$.

The main result of this work is that the ground state degeneracy $D$ of $H$ is encoded in $|\tilde{\psi}\rangle$, the ground state of $\tilde{H}$ obtained by evolution-based method starting from the vacuum states of $N$ ququarts $|\tilde{\bf0}\rangle=\bigotimes^N|\tilde{0}\rangle$, with $|\tilde0\rangle$ the vacuum state of a single ququart, i.e.
\begin{equation}
    |\tilde{\psi}\rangle\propto \lim_{\tau\rightarrow\infty}f(\tau,\tilde{H})|\tilde{\mathbf 0}\rangle.\label{eq:evolution}
\end{equation}
This evolution reduces to the nominal imaginary time evolution when $f(\tau, \tilde{H})$ is chosen as $e^{-\tau \tilde{H}}$.

We will first illustrate how degeneracy $D$ is encoded in $|\tilde{\psi}\rangle$ and discuss the different decoders for changing situations in the following.
A local Hilbert space $|\tilde\alpha\rangle$ with $\alpha=0,1,2,3$ can be formed by mapping $O_\alpha$ into $|\tilde\alpha\rangle$, which is equivalent to a unitary transformation defined by a $4\times 4$ matrix $\mathcal{O}$ with matrix elements ${(\mathcal{O})}_{\alpha,pq} \triangleq {O_\alpha^{pq}}$.
A Hamiltonian in the whole Hilbert space $H'$ is unitary transformed to a state in ququart space $|H'\rangle$ by $\mathcal{O}^{\otimes N}$ and denoted by $\sim$ in the following,
\begin{equation}
    \sum\limits_{\{\alpha\}} c_{\{\alpha\}}' \bigotimes_{k=1}^{N} O_{\alpha_k}=H'\sim|H'\rangle = \sum\limits_{\{\alpha\}} c_{\{\alpha\}}' \bigotimes_{k=1}^{N} |\tilde{\alpha_k}\rangle.\label{eq:defstate}
\end{equation}
A direct calculation then gives\begin{equation}
    \langle H'|H\rangle = \sum\limits_{\{\alpha\}\{\alpha'\}}c_{\{\alpha'\}}'^*c_{\{\alpha\}}\delta_{\{\alpha'\}\{\alpha\}}=\Tr(H'^\dag H),\label{eq:innerprod}
\end{equation}\vspace*{-0.5cm}
\begin{equation}
    \sum\limits_{\{\alpha\}} c_{\{\alpha\}} \bigotimes_{k=1}^{N} \tilde{O}_{\alpha_k}|H'\rangle\triangleq\tilde{H}|H'\rangle=|HH'\rangle.\label{eq:HH}
\end{equation}
Similarly, it is also valid to define $\tilde{O}^{\beta\gamma}_\alpha$ as $\Tr(O_\beta^\dag O_\gamma O_\alpha)$ to ensure $\tilde{H}|H'\rangle$ equivalent to $|H'H\rangle$. We adopt the average of two $\tilde{H}$ definitions for numerical convenience in this work.
The detailed proof and an alternative description for the above formula in tensor language can be found in the Supplementary Material (SM).

The complexity of finding a ground state of a Hamiltonian depends on the structure of the Hamiltonian and their spectra details.
The two Hamiltonians $\tilde{H}$ and $H$ share the same coefficients $c_{\{\alpha\}}$, thus exhibit the same structure.
The commutation relationship of $\tilde{O}_\alpha$ can be shown to be the same as $O_\alpha$~\cite{sm},\begin{equation}
    \begin{aligned}
    \relax[O_\alpha,O_\beta] &= \sum\limits_\gamma (\tilde{O}_\alpha^{\gamma\beta}-\tilde{O}_\beta^{\gamma\alpha})O_\gamma,\\
    [\tilde{O_\alpha},\tilde{O_\beta}] &= \sum\limits_\gamma (\tilde{O}_\alpha^{\gamma\beta}-\tilde{O}_\beta^{\gamma\alpha})\tilde{O_\gamma}.
    \end{aligned}
\end{equation}
The spectra of the two Hamiltonians are also identical and there is a $2^N$ to $1$ correspondence between their eigenstates~\cite{sm}.
Thus, the complexities of solving ground states of $\tilde{H}$ and $H$ are at a similar level, although one of the ground states of $\tilde{H}$ encodes the solution to a harder \#BQP problem.

As the first Pauli basis $O_0$ is proportional to the identity matrix, the normalized state $|\tilde{\mathbf{0}}\rangle$ equals to $2^{-N/2}|I\rangle$ according to Eq.~(\ref{eq:defstate}), with $I$ the identity Hamiltonian of $N$ qubits.
The evolution of the super Hamiltonian can then be regarded as
\begin{equation}
    |\tilde{\psi}\rangle \propto \lim_{\tau\rightarrow\infty}f(\tau,\tilde{H})|\tilde{\mathbf{0}}\rangle = 2^{-N/2}\left|\lim_{\tau\rightarrow\infty}f(\tau,H)\right\rangle,
\end{equation}
according to Eq.~(\ref{eq:innerprod}).
When $\tau$ approach infinity, $f(\tau, H)$ is proportional to the projector of the ground subspace $P_{\rm gs}$ with equal weights on all degenerate ground components.
Thus, $|\tilde\psi\rangle \sim 1/\sqrt{D} P_{\rm gs}$ with the prefactor determined by the normalization of $|\tilde\psi\rangle$, which naturally encodes the information of the ground state degeneracy.

\paragraph*{Phase boundary of 1D system.}

For 1D system, MPS is a powerful ansatz for both accuracy and efficiency.
We will illustrate the above algorithm with the MPS form in the following.
The Trotter expansion in the standard imaginary time evolution~\cite{Daley2004} will involve extra errors unless the time step chosen is small enough.
Thus, we present the results from the power method instead, as it is found to be more convenient and efficient in our numerical calculations.
We first shift the Hamiltonian by $e_0$ to assure the ground state has the largest absolute energy (a practical approach is discussed in the SM).
With $f(\tau, \tilde{H}) = {(\tilde{H}-e_0)}^\tau$ in Eq.~(\ref{eq:evolution}), we find \begin{equation}\label{equ:degeneracy}
    \langle\tilde{\mathbf{0}}|\tilde\psi\rangle = 2^{-N/2} D^{-1/2} \Tr(I^\dag P_{\rm gs})=2^{-N/2} D^{1/2}.
\end{equation}
The exponentially small factor is caused by the normalization of Pauli basis $O_0$, which can be avoided by absorbing a factor $\sqrt{2}$ into the MPS tensor of $|\tilde{\mathbf{0}}\rangle$ at each site.

\begin{figure}[htb]
    \centering
    \inputtikz{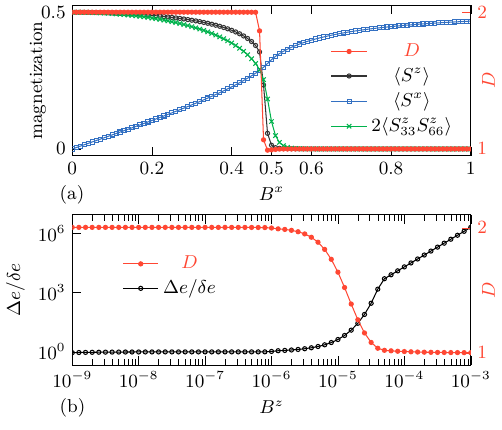}
    \caption{(a)~The magnetization and degeneracy of a $100$-site ferromagnetic transverse field Ising model with a longitudinal field $B^x$. A pinning field $B^z=1\times10^{-5}$ is applied when calculating $\langle S^z\rangle$.\ (b)~The degeneracy and $\Delta e/\delta e$ at $B^z$ with $\Delta e$ the energy gap introduced by $B^z$ and $\delta e$ the numerical energy uncertainty of the final state. The virtual bond dimension chosen is $60$ for all simulations.\label{fig:ising}}
\end{figure}

We first consider the ferromagnetic transverse field Ising model with Hamiltonian in the form\begin{equation}
    H_{\rm TFI} = -\sum S_i^z S^z_{i+1} + B^x \sum S_i^x +B^z \sum S_i^z.
\end{equation}
When $B^z=0$, this model exhibits two phases divided by the critical field $B^x_{\rm c}=0.5$: a ferromagnetic phase with two degenerate ground states for smaller $B^x$ and a paramagnetic phase with a unique ground state for larger $B^x$.
Magnetization and the long-range magnetization correlation are commonly adopted to describe this phase transition.
They are obtained by variational MPS methods with bond dimension equal to $60$ and graphed in Fig.~\ref{fig:ising}(a).
To mitigate the magnetization cancellation arising from two degenerate states when calculating $\langle S^z\rangle$, a tiny pinning field $B^z=1\times10^{-5}$ is introduced.
To eliminate boundary effects, the magnetization correlation is computed by considering the 33rd and 66th sites out of the total 100 sites.
The degeneracy calculated with the same bond dimension is also illustrated.
The value of degeneracy is always an integer, thus it gives the sharpest transition edge and the most accurate phase boundary.

It is usually difficult to determine whether two numbers are exactly the same or not numerically due to the finite numerical precision or round off errors.
We fix $B^x=0$ and add a tiny gap $\Delta e$ to separate the degenerate ground states by applying a small magnetic field $B^z$.
Degeneracy calculated from a stable algorithm will be unaffected by a sufficiently small $B^z$, but should reduce to $1$ for sufficiently large $B^z$.
The calculated degeneracy together with the $\Delta e/\delta e$ is illustrated in Fig.~\ref{fig:ising}(b), with $\delta e$ the energy uncertainty of the final state.
At small $B^z$, the energy gap is comparable to the energy uncertainty thus our algorithm cannot tell the difference between the two neighboring states and gives $D=2$.
The turning point for this change of degeneracy and $\Delta e/\delta e$ coincides nicely.
For evolution-based methods, a smaller energy uncertainty is expected for longer evolution time. This provides a systematic way to achieve better energy resolution as well as the degeneracy resolution.

\begin{figure}[htb]
    \centering
    \inputtikz{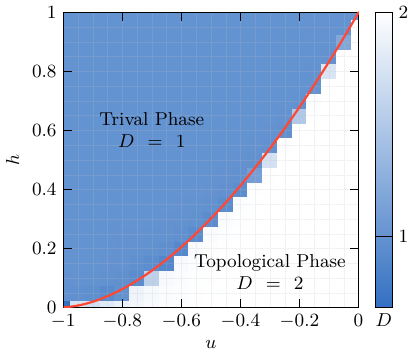}
    \caption{Phase diagram of a 100-site Kitaev-Hubbard model. The topological phase boundary is marked by the calculated degeneracy at bond dimension $60$.\label{fig:kitaev}}
\end{figure}

Topological phase is identified by the existence of zero mode, which is closely related to the ground state degeneracy. While a basic understanding of zero mode comes from non-interacting systems, their interacting counterparts, the weak zero mode, also attracts much attention~\cite{Alicea2016}.
Carrying out these studies are difficult because no general method is known for the detection of topological zero mode in an interacting system.
By adopting our algorithm, all evolution-based methods can be used to detect such zero mode in an interacting system through direct numerical studies.
As a benchmark for our algorithm, we introduce the Kitaev-Hubbard model
\begin{equation}
    \begin{aligned}
        H_{\rm KH} =& -\sum(c_i^\dag-c_j)(c^\dag_{j+1}+c_{j+1}) - h\sum(1-2c_j^\dag c_j)\\
        &+u\sum(1-2c_j^\dag c_j)(1-2c_{j+1}^\dag c_{j+1}),
    \end{aligned}
\end{equation}
whose topological phase boundary~\cite{Mahyaeh2020} is shown by the red line in Fig.~\ref{fig:kitaev}.
We sweep all points in the phase diagram and calculate their corresponding degeneracies as shown by the background colors.
The two-fold degeneracy means the existence of a weak zero mode and correspondingly a topological phase.
Our results are found to be in excellent agreement with the known topological phase boundary.

\paragraph*{Lanczos method for 2D frustrated system.}

Lanczos method can iteratively solve for several extreme eigenvalues and eigenvectors of a Hermitian system without requiring explicit storage of the complete Hamiltonian $H$.
It simplifies the problem with the Krylov subspace, which is spanned by the vectors $H^i\vec{v}_0$ with $\vec{v}_0$ an initial vector and $i$ the power exponent. All the vectors involved in the Lanczos method take the form of $\vec{v}_0$ multiplied by a polynomial of $H$ of order $\tau$, noted as ${\rm Ploy}(\tau, H) \vec{v}_0$.
During the iteration process, the order of the polynomial $\tau$ increases until converge is reached.
Thus, the Lanczos method can be recast in the form of evolution of operators,
\begin{equation}
    |{\rm gs}\rangle=\lim\limits_{\tau\rightarrow\infty} f(\tau, H)|x\rangle = \lim\limits_{\tau\rightarrow\infty}{\rm Ploy}(\tau, H)|x\rangle.
\end{equation}
After getting $|\tilde\psi\rangle$ as a dense vector $\tilde{\vec{v}}$, the degeneracy can be calculated analogously to Eq.~(\ref{equ:degeneracy}) with the overlap $\langle \tilde{\mathbf{0}}|\tilde{\psi}\rangle$ simplified as the first element of this vector.

While some may consider employing the conventional Lanczos method to extract the lowest few eigenvalues and determine degeneracy, this approach faces the challenge of excited states collapsing into lower-energy states.
Furthermore, it can be intricate to establish a criterion for determining when two energy levels are sufficiently close to be treated as degenerate.
In practice, we also find that degenerate eigenvalues can sometimes be missed when applying the conventional Lanczos method with a random initial state, especially when the transverse field $B^x$ is small.
However, our algorithm circumvents such difficulties by requiring only one ground state to obtain the degeneracy information.

If a Hamiltonian possesses conserved quantities other than energy and the initial state $\vec{v}_0$ is in a different conserved sector from the ground states, the Krylov space spanned by $\vec{v}_0$ cannot cover the ground states and the Lanczos method will be stuck at a local solution.
In practical implementations, a restart scheme is adopted to avoid such impasse.
When the spanned Krylov space is invariant after further increasing of the polynomial order $\tau$, a random vector $\vec{v}_0'$ is selected to restart the growth of Krylov space.
In our algorithm, the initial state is chosen to represent the identity Hamiltonian which represents a combination of all conserved sectors, thus there is no need to adopt the restart scheme.
Besides, the restart scheme will erase information of degeneracy.
More details of our implementation can be found in the SM\@.

\begin{figure}[htb]
    \centering
    \inputtikz{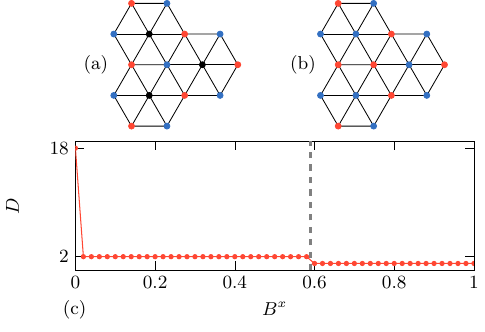}
    \caption{(a)(b)~the ground state configuration on triangle lattice in the absence of transverse field $B^x$. The red and blue dots denote two different spin components. The black dots indicate frustrated sites and both spin components minimize the system energy.\ (c)~The dependence of degeneracy on the transverse field $B^x$. The frustration disappears at a finite $B^x$ and the $\rm Z_2$ symmetry is broken at the critical field $B^x_{\rm c}=0.59$ (vertical dashed line).\label{fig:2d}}
\end{figure}

We illustrate our algorithm by a frustrated transverse field Ising model on a triangular lattice.
When the transverse field is zero, tropical tensor network~\cite{Liu2021} can be used to study the degeneracy of this system analytically.
However, the transverse field $B^x$ introduces fluctuation between different degenerate states which cannot be properly handled by the former method.
We consider the configuration constituted by three hexagons as illustrated in Fig.~\ref{fig:2d}(a).
At zero field, to minimize energy, all the edges of the three hexagons form an antiferromagnetic pattern and are colored in red or blue.
The three centers of hexagons are frustrated and give $2^3=8$ configurations.
A different pattern (Fig.~\ref{fig:2d}(b)) happens to have the same energy giving a total of $9$ configurations.
After considering $\rm Z_2$ symmetry, the overall degeneracy becomes $2\times9=18$.
In this model, such frustration is very fragile, and will be destroyed by a tiny transverse field.
As shown in Fig.~\ref{fig:2d}(c), the degeneracy is reduced to $2$ from $18$ with a non-zero transverse field $B^x$, which further reduces to the trivial degeneracy due to the breaking of $\rm Z_2$ symmetry.
The critical field is clearly shown to be $0.59$ in the figure.

\paragraph*{3D quantum chemistry system.}

Quantum chemistry systems are believed to represent a promising application scenario to show quantum advantage.
Imaginary time evolution is a standard method to solve such systems by quantum circuits, such as QITE~\cite{McArdle2019,Motta2019}.
This extends our algorithm by the quantum circuits.
However, Eq.~(\ref{equ:degeneracy}) cannot be applied directly due to its exponentially small prefactor.
We adopt another scheme based on overlap estimation——after the procedure of QITE on the super Hamiltonian, a QITE is performed to get one of the ground states $|\phi\rangle$ of the original Hamiltonian $H$.
Then one can obtain\footnote{We assume all states are real here} \begin{equation}
    (\langle\phi|\otimes\langle\phi|)P\mathcal{O}^{\otimes N}|\tilde{\psi}\rangle=1/\sqrt{D}\langle\phi|P_{\rm gs}|\phi\rangle = 1/\sqrt{D}.\label{eq:qite}
\end{equation} Here, $P$ denotes a permutation making each $\mathcal{O}$ act on the ququart/qubit of $|\tilde{\psi}\rangle$/$\langle\phi|$ with the same rank from $1$ to $N$, and $P\mathcal{O}^{\otimes N}$ transforms $|\tilde{\psi}\rangle$ back to $1/\sqrt{D}P_{\rm gs}$.
The left-hand side can be measured by a swap test~\cite{Barenco1997} or its variant~\cite{Fanizza2020}.

\begin{figure}[htbp]
    \centering
    \inputtikz{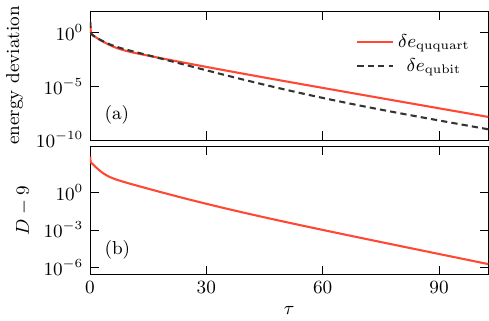}
    \caption{(a)~energy deviation $\delta e$ and (b)~degeneracy $D$ with respect to the imaginary time $\tau$.\label{fig:qite}}
\end{figure}

We adopted an ab-initio Hamiltonian of a single carbon atom with basis set 6-31g and a total of $10$ qubits are needed for both spin components of the original Hamiltonian and $20$ qubits for the super Hamiltonian.
The Hamiltonian contains the kinetic energy and Coulomb repulsion without interaction between different spins.
Thus, no Hund's rules apply in this minimal model and the three degenerated 2p orbitals possess only two electrons for each spin component leading to a total degeneracy $D=3\times3=9$.
To corroborate the theoretical validity of the aforementioned scheme, we performed numerical simulations of the whole QITE process, and the results are illustrated in Fig.~\ref{fig:qite}.

For system with finite degeneracy, the overlap to estimate is a finite number, which can be measured efficiently by overlap estimation.
In Fig.~\ref{fig:qite}(a), the converge rate of $\tilde{H}$ is found to be slightly slower than that of $H$.
As the spectra is identically the same between the super Hamiltonian $\tilde{H}$ and the original Hamiltonian $H$, the different initial states account for their differences.
For $H$, a random initial state is adopted while only the $|\tilde{\mathbf{0}}\rangle$ can be adopted for $\tilde{H}$.
In more complicated systems, such as strongly correlated systems, the tail of the converge curve will be long and can dominate the whole process.
The influence of the initial states would be much smaller.
Our algorithm is shown to be capable of solving the \#BQP problem with an analogous QITE procedure for solving a QMA problem, which suggests that our algorithm represents a better approach to achieve quantum acceleration over classic computation, especially in the current stage where the larger number of qubits is more accessible compared to deeper circuits.

However, for systems with exponentially large degeneracy, such as spin glass models, the overlap to estimate is exponentially small. The steps needed in estimating the overlap will also be exponentially large which will in turn require exponentially deep circuits. For the time being, such systems are beyond the capacity of the quantum circuits implementation of our algorithm.

\paragraph*{Discussions.}

We present an approach and demonstrate numerically its possibility to solve the \#BQP problem with evolution-based method normally used to solve for QMA problems.
Our work does \bf not \mdseries imply that the two classes \#BQP and QMA are equivalent, though.
As an analogy, we note, the complexity of solving for ground state of a non-interacting system and a strongly correlated system by exact diagonalization are the same, but difficulties for solving these two systems are quite different.

Comparing to solving for ground states by evolution-based methods problem, our algorithm doubles the system size while maintaining the same complexity scaling.
The choice of initial state is restricted, and some approximate initial state can not be applied to accelerate the evolution.
The requirement of the accuracy of the evolution is also higher.
For instance, the usually adopted imaginary time evolution can take large time steps in the beginning and decrease gradually to accelerate the process, which will lead to large error in our algorithm.
Furthermore, if the quantum states are represented by actual qubits instead of numerical data, the resources needed to decode the results should also be considered.
To decode an exponentially large degeneracy in sufficient accuracy is much harder than to decode expectation energy of a quantum state.
However, our algorithm essentially solves the \#BQP problem, which is believed to be in a harder complexity class than the problem of finding the ground states, even when P is equal to NP.
Our algorithm demonstrates that from the perspective of evolution-based methods, the QMA problem is equivalent  to finding the ground states of a $k$-local Hamiltonian, while \#BQP corresponds to identifying a special ground state with twice the number of qubits.
The additional complexity inherent in identifying the specific ground state is encapsulated in the aforementioned statement.

The key requirement to the evolution in this work is that it should remain unbiased to the different degenerate ground state components.
Besides the methods mentioned in the main text, the well known simple/full update~\cite{Jiang2008, Lubasch2014}, Monte Carlo gradient methods for projected entangled-pair states~\cite{Liu2017} are also unbiased in principle and can be combined with our algorithm.
Furthermore, combining our algorithm with the tangent space method enables its extension to infinite systems.

\paragraph*{Conclusion.}

The main result of this work concerns the proposed super Hamiltonian and the proof of its relationship to the original Hamiltonian.
It is shown possible to obtain the ground state degeneracy, a \#BQP problem, by directly simulating the evolution of operators with existing methods used to solve for the ground state, a QMA problem.
Our work thus provides a better understanding of the connections between these two complexity classes and helps to extend the existing methods to harder problems.
For applications, by combining with various methods for different systems and dimensions, our algorithm is shown to constitute an effective approach for describing phase transitions, detecting topological zero modes, and studying frustrated systems, etc.

\begin{acknowledgments}
    We thank Yang Suo, Meng Li, Yifei Huang, Changsu Cao, Dingshun Lv, and Bo Zhan for their help and discussions. This work is supported by  the National Natural Science Foundation of China (NSFC) (Grants Nos. 92265205, 12361131576, and 12174214) and by Innovation Program for Quantum Science and Technology (2021ZD0302100).
\end{acknowledgments}

\bibliography{Ref}

\clearpage
\section*{Appendix}

\subsection*{A brief introduction to different complexity classes}
\begin{figure}[htbp]
    \inputtikz{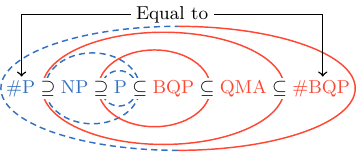}
    \caption{The relationship of complexity classes involved in this work. Blue (red) is used to represent the classic (quantum) one.\label{fig:PNP}}
\end{figure}

The complexity class of nondeterministic polynomial time (NP) holds a significant position in computational theory, attributable to its fundamental application across a variety of domains.
An NP problem is a decision problem, which determines whether a solution satisfying a certain condition exists.
On the other hand, a \#P problem extends beyond decision-making, requiring the enumeration of the total number of valid solutions.
The complexity of \#P problems is generally considered superseding that of NP problems even in situation where P is equal to NP\@.
Quantum computing, enhanced by quantum mechanics, is perceived to possess greater computational power than its classical counterparts.
Certain NP problems can be solved within polynomial time using quantum computing, thereby defining the class of bounded-error quantum polynomial time (BQP) problems.
However, quantum computing does not extend its advantage to resolving \#P problems: \#BQP problem is surprisingly proven to be equivalent to its classical analog~\cite{Brown2011}, \#P (as illustrated in Fig.~\ref{fig:PNP}).
Within the framework of Hamiltonians, quantum Merlin Arthur (QMA) problem, the quantum version of NP problem, aimed at solving for the ground state energy of a $k$-local Hamiltonian for $k\ge2$, and the \#BQP problem need to further count the degeneracy of these ground states.
The equivalence of \#P and \#BQP underscores a noteworthy equivalence in complexity between counting the ground state degeneracy of quantum and classical Hamiltonians.

\subsection*{Proof of Eq.~(\ref*{eq:HH})}

By definition\begin{equation}
    {(\tilde{O}_\alpha^{\beta\gamma})}^{(L)} \triangleq \Tr (O_\beta^\dag O_\alpha O_\gamma),
\end{equation}
where $O_\alpha O_\gamma$ gives a new operator in qubit space thus can also be decomposed by the same basis $B$, $O_\alpha O_\gamma = \sum\limits_\beta c_{\alpha\beta\gamma} O_\beta$.
By left-multiplying $O_\beta^\dag$ and using the orthonormal condition, we can obtain the coefficient $c_{\alpha\beta\gamma} = \Tr(O_\beta^\dag O_\alpha O_\gamma)={(\tilde{O}_\alpha^{\beta\gamma})}^{(L)}$.
Then we can obtain
\begin{equation}
    \begin{aligned}
        H H' &= \sum\limits_{\{\alpha,\gamma\}}c_{\{\alpha\}}c_{\{\gamma\}}'\bigotimes_{k=1}^{N} O_{\alpha_k}O_{\gamma_k} \\
        &= \sum\limits_{\{\alpha,\gamma\}}c_{\{\alpha\}}c_{\{\gamma\}}'\bigotimes_{k=1}^{N} \sum\limits_{\beta_k}{(\tilde{O}_{\alpha_k}^{\beta_k \gamma_k})}^{(L)} O_{\beta_k}\\
        &\sim\sum\limits_{\{\alpha,\gamma\}}c_{\{\alpha\}}c_{\{\gamma\}}'\bigotimes_{k=1}^{N} \sum\limits_{\beta_k}{(\tilde{O}_{\alpha_k}^{\beta_k \gamma_k})}^{(L)}|{\tilde{\beta}_k}\rangle\\
        &=\sum\limits_{\{\alpha,\gamma\}}c_{\{\alpha\}}c_{\{\gamma\}}'\bigotimes_{k=1}^{N} \tilde{O}_{\alpha_{k}}^{(L)}|{\tilde{\gamma}_{k}}\rangle\\
        &=\tilde{H}^L|H'\rangle,
    \end{aligned}
\end{equation}

\subsection*{Obtain super Hamiltonian in the language of tensor}

\begin{figure}[htbp]
    \inputtikz{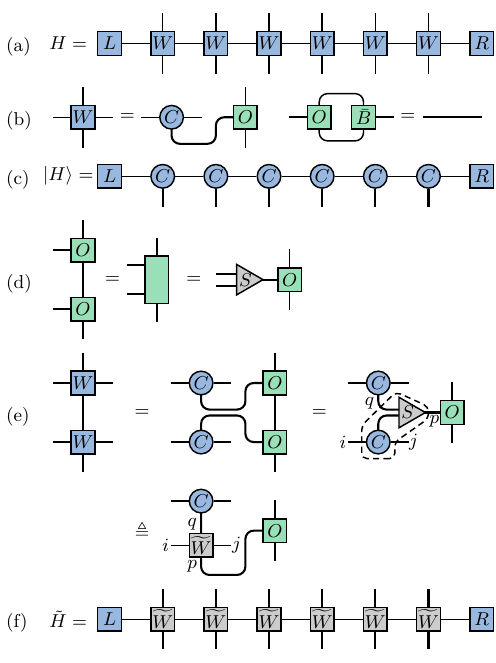}
    \caption{(a) A Hamiltonian in MPO form.\ (b) Decomposition of the MPO tensor $W$ into basis tensor $B$ and coefficient tensor $C$. $\bar{B}$ denotes conjugation of $B$.\ (c) A MPS formed by coefficient tensors represents the Hamiltonian.\ (d) Multiplication of two basis tensors.\ (e) Multiplication of two MPOs.\ (f) The new Hamiltonian $\widetilde{H}$ associated with $H$. The thicker bonds represent dimension identical to the square of physical dimensions.\label{fig:tensor}}
\end{figure}

In most MPS-based methods, $H$ is represented by matrix product operator (MPO), as illustrated in Fig.~\ref{fig:tensor}(a).
A rank-4 tensor $W_{ij}^{pq}$ sits on a site with two vertical physical indices $p$ and $q$, and two horizontal virtual indices $i$ and $j$.
Given $i$ and $j$, $W$ reduces to a linear operator on a single site and can be expanded on a set of basis indexed by $\alpha $ under the Hilbert-Schmidt inner product $X \cdot Y=\Tr(X^\dag Y)$.

The Pauli basis is a natural choice for spin-$1/2$ system.
For larger spins, Gell-Mann matrices can be used instead.
All these matrices, except for the identity matrix, are Hermitian and traceless.
These two properties are not necessarily required for our algorithm, but they will become convenient for numerical simulations.
If the basis is Hermitian, real expansion coefficients correspond to Hermitian Hamiltonians and vice verse.
If all basis (except the first one) are traceless, the trace of the operator equals the amplitude of the $|H\rangle $ when all physical indices are chosen as the first one.
For more complicated systems, other basis can also be employed for our algorithm as long as they are complete and orthonormal, including the numerical basis obtained by QR decomposition of the MPO tensor $W_{ij}^{pq}$ ($pq$ being the physical indices and $ij$ being the virtual ones).

As shown in Fig.~\ref{fig:tensor}(b), this process can be regarded as decomposition of $W_{ij}^{pq}$ onto a set of basis $O_\alpha$,
\begin{equation}
    W_{ij}^{pq} = \sum C_{ij}^\alpha O_\alpha^{pq},\label{eq:decom}
\end{equation}
where the basis tensor $O_\alpha^{pq}$ is formed by a set of orthonormal basis, i.e. $\sum {(O_\alpha^{pq})}^* O_\beta^{pq}=\delta_{\alpha\beta}$, and the coefficient tensor $C_{ij}^\alpha $ can be obtained by contracting $W$ and the conjunction of $O$,
\begin{equation}
    C_{ij} ^\alpha=\sum {(O_\alpha^{pq})}^*W_{ij}^{pq}.\label{eq:hatting}
\end{equation}
The slight thicker legs are employed to symbolically represent the bonds with dimension identical to the square of the physical dimension.
All information of $W$ is contained in coefficient tensor $C$ which is in a one-to-one correspondence with an MPO tensor $W$.
In this way, the Hamiltonian $H$ can be represented by a matrix product state $|H\rangle$ formed by coefficient tensor $C$, as illustrated in Fig.~\ref{fig:tensor}(c).
We use the notation $H\sim|H\rangle $ to emphasize that two physical indices in the MPO form of $H$ are fused into one basis index in the MPS form of $|H\rangle $. $H$ and $|H\rangle$ are connected by a unitary transformation defined by $\mathcal{O}^{\otimes N} $.

The decomposition in Eq.~(\ref{eq:decom}) separates the virtual and the physical indices into the coefficient tensor or the basis tensor respectively.
Because the multiplication of two Hamiltonians in MPO form is the contraction of physical indices, only basis tensors are involved in the multiplication.
Besides, the multiplication of two basis tensors can also be decomposed according to the same basis, i.e.
\begin{equation}
    \sum O_\alpha^{pq} O_{\beta}^{qr} = \sum S_{\alpha\beta}^\gamma O_\gamma^{pr},\label{}
\end{equation}
as illustrated in Fig.~\ref{fig:tensor}(d).
The structure tensor $O$ describes the multiplication relationship of the basis and is independent of the specific Hamiltonian.
The multiplication of two MPO tensors can be represented by the contraction of two coefficient tensors, a structure tensor and a basis tensor (Fig.~\ref{fig:tensor}(e)),
\begin{equation}
    \begin{aligned}
        \sum\limits W_{2,ij}^{pq}W_{1,kl}^{qr}&=\sum C_{2,ij}^\alpha C_{1,kl}^\beta S_{\alpha\beta}^\gamma O_\gamma^{pr}\\&\triangleq \sum \widetilde{W}_{2,ij}^{\gamma\beta}C_{1,kl}^{\beta}O_\gamma^{pr},\label{eq:mpo}
    \end{aligned}
\end{equation}
where $\widetilde{W}_2$ forms an MPO representation of a super Hamiltonian $\widetilde{H}_2$ (Fig.~\ref{fig:tensor}(f)) acting on $|\psi(H_1)\rangle $.
The multiplication of two Hamiltonians $H_2\times H_1$ is converted into an action of $\widetilde{H}_2$ on a state $|\psi(H_1\rangle)$, i.e., $H_2\times H_1\succ \widetilde{H}_2|\psi(H_1)\rangle$.

We emphasize that $\widetilde{W}_2$ and $W_2$ have the same coefficient tensor.
If we treat $S_{\alpha\beta}^\gamma$ as a new basis with $\alpha$ the basis index and $\gamma,\beta$ the physical indices, i.e. ${(\tilde{O}_\alpha^{\gamma\beta})}^{(L)}$. $\widetilde{W}_2$ is a basis replacement to $W_2$.

\subsection*{Proof of the identical commutation relation of the two basis}

The commutation relation of basis $O$
\begin{equation}
    \begin{aligned}
        \relax[O_\alpha, O_\beta] & = \sum O_\alpha^{pq} O_\beta^{qr} - O_\beta^{pq} O_\alpha^{qr} \\
        & = \sum (S_{\alpha\beta}^\gamma-S_{\beta\alpha}^\gamma) O_\gamma^{pr}
    \end{aligned}
\end{equation}
is contained in the tensor $S_{\alpha\beta}^\gamma-S_{\beta\alpha}^\gamma$.
The commutation relation of basis $S$ reads
\begin{equation}
    \begin{aligned}
        \relax[S_\alpha, S_\beta] & = \sum S_{\alpha\lambda}^\mu S_{\beta\nu}^\lambda - S_{\beta\lambda}^\mu S_{\alpha\nu}^\lambda\\
        & = \sum S_{\alpha\beta}^\gamma S_{\gamma\nu}^\mu - S_{\beta\alpha}^\gamma S_{\gamma\nu}^\mu\\
        & = \sum (S_{\alpha\beta}^\gamma-S_{\beta\alpha}^\gamma) S_{\gamma\nu}^\mu,\label{eq:commute}
    \end{aligned}
\end{equation}
which is the same as for $O$.
The second step in Eq.~(\ref{eq:commute}) uses the identity $S_{\alpha\lambda}^\mu S_{\beta\nu}^\lambda = S_{\alpha\beta}^\gamma S_{\gamma\nu}^\mu$, which comes from the associative law as proven in the following.

\begin{figure}[ht]
    \centering
    \hspace*{-0.5cm}\inputtikz{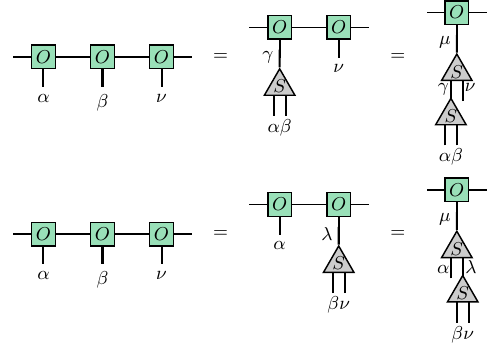}
    \caption{Associative law of basis multiplication.\label{fig:associativity}}
\end{figure}

According to the associative law, different calculation orders for the multiplication of three basis tensors will give the same answer, as in the following equations
\begin{equation}
    \begin{aligned}
    &\sum O_\alpha^{pq}O_\beta^{qr}O_\nu^{rs} = \sum (O_\alpha^{pq}O_\beta^{qr})O_\nu^{rs} \\
    =& \sum S^\gamma_{\alpha\beta}O_\gamma^{pr}O_\nu^{rs} = \sum S^\gamma_{\alpha\beta} S^\mu_{\gamma\nu} O_\mu^{ps},\label{equ:m31}
    \end{aligned}
\end{equation}
\begin{equation}
    \begin{aligned}
    &\sum O_\alpha^{pq}O_\beta^{qr}O_\nu^{rs} = \sum O_\alpha^{pq}(O_\beta^{qr}O_\nu^{rs}) \\
    =& \sum O_\alpha^{pq} S^\lambda_{\beta\nu}O_\lambda^{qs} = \sum S^\lambda_{\beta\nu} S^\mu_{\alpha\lambda} O_\mu^{ps},\label{equ:m32}
    \end{aligned}
\end{equation}
with Fig.~\ref{fig:associativity} a pictorial illustration.

The equality of Eqs.~(\ref{equ:m31}) and (\ref{equ:m32}) gives \begin{equation}S_{\alpha\lambda}^\mu S_{\beta\nu}^\lambda = S_{\alpha\beta}^\gamma S_{\gamma\nu}^\mu.\end{equation}

\subsection*{Relationship of eigenvalues and eigenvectors between a Hamiltonian and its super Hamiltonian}

The original Hamiltonian $H$ can be decomposed into its eigenbasis as
\begin{equation}
    H=\sum e_i|\phi_{ij}\rangle \langle \phi_{ij}|,
\end{equation}
with $i$ the energy index, and $j$ the degeneracy index.
If $|\tilde{\psi}\rangle$ is an eigenstate of the super Hamiltonian $\widetilde{H}$, i.e.
\begin{equation}
    \widetilde{H}|\psi\rangle = \varepsilon |\psi\rangle,
\end{equation}
with $\varepsilon $ the eigenenergy, an operator $P(\psi) \sim |\psi\rangle$ can be found by transforming back with ${\mathcal{O}}^{\otimes N} $ such that
\begin{equation}
    HP(\psi)=\varepsilon P(\psi).\label{eq:eigen}
\end{equation}
$P(\psi)$ is an operator in the original Hilbert space, which can also be decomposed as
\begin{equation}
    P(\psi) = \sum c_{ijkl}|\phi_{ij}\rangle\langle\phi_{kl}|.\label{eq:decomp}
\end{equation}
Equations (\ref{eq:eigen}) and (\ref{eq:decomp}) give
\begin{equation}
    \sum e_i c_{ijkl}|\phi_{ij}\rangle\langle\phi_{kl}|=\varepsilon \sum c_{ijkl}|\phi_{ij}\rangle\langle\phi_{kl}|,
\end{equation}
which indicates $\varepsilon $ is also one of the eigenenergy of $H$, $\varepsilon = e_p$ for example, and $c_{ijkl} = 0$ if $ i \ne p$.

On the other hand, for given eigenstates $|\phi_{ij}\rangle$ of $H$, one can construct $\mathcal{N}$ linear independent operators
\begin{equation}
    P_{kl} = |\phi_{ij}\rangle\langle \phi_{kl}|
\end{equation}
with $\mathcal{N}$ the dimension of Hilbert space such that
\begin{equation}
    HP_{kl} = e_i P_{kl},
\end{equation}together with $\mathcal{N}$ linear independent $|\psi_{kl}\rangle \sim P_{kl}$ satisfying
\begin{equation}
    \widetilde{H}|\psi_{kl}\rangle = e_i|\psi_{kl}\rangle.
\end{equation}

\subsection*{How to shift the Hamiltonian}
The choice of the constant shift $e_0$ is not unique.
A too small $e_0$ will not be capable of ensuring the ground states have the largest absolute energy, thus leading to a wrong degeneracy.
A too large $e_0$ will decrease the ratio between the energies of the ground and the first excited states thus making the power iteration less efficient.
Based on our experience, the summation of the max energies of all local terms in the Hamiltonian usually represents an acceptable choice.
And it turns out that this is indeed the maximum energy for a frustration-free Hamiltonian.

\subsection*{Detail of Lanczos methods}
\begin{algorithm}[htb]
    \caption{Lanczos method\label{alg:Lanczos}}
    \DontPrintSemicolon
    \SetKwInOut{Input}{input}
    \SetKwInOut{Output}{output}
    \Input{Hamiltonian Function: {\rm FuncH},\linebreak
    Initial vector: $V_0$,\linebreak
    Number of Krylov space to keep: ndim,\linebreak
    Maximum iteration times: maxiter.}
    \Output{$e$, $V$}

    nv = length($V_0$)

    $V_1 = V_0$

    Initialize nv$\times$ndim basis matrix $V_s$ with zeros

    \For{${\rm count} \in 1\sim \rm maxiter$}{

    $V_2 = {\rm FuncH}(V_1)$

    Project $V_2$ to the subspace $I-V_s$

    \eIf{{\rm count > ndim}}{
        Local Hamiltonian $h=V_s^\dag {\rm FuncH}(V_s)$

        $e, v = {\rm Eig}(h)$ with $e$ in ascending order

        \If{$e[0]$ \rm converged}{\Return $e[0], V_s\times v[:, 0]$}

        $V_s = V_s \times v$

        Replace last row of $V_s$ with $V_2$

    }{
        $V_s[:, {\rm count}] = V_2$
    }
    $V_1=V_2$
    }
    \Return $e[0], V_s\times v[:, 0]$
\end{algorithm}

From an initial vector $V_0$, a Krylov subspace is constructed first as $\mathcal{L}_{i=0}^n(H^i V_0)$, where $\mathcal{L}(\cdot)$ means the linear space spanned by listed vectors and $n$ is chosen as large as possible until the limitation of computation resources is reached.
In this subspace a set of orthogonal basis is obtained by Gram-Schmidt process and is stored as the rows of matrix $V_s$.
A diagonalization is then performed in the subspace Hamiltonian $h=V_s^\dag H V_s$ and the subspace is shrunk by projecting out the basis with the highest energy.
The remaining $n-1$ basis together with the $H V$ span the new Krylov subspace.
In the whole procedure, all vectors remain in the form like $\rm{Ploy}(\beta, H)V_0$.
The detailed algorithm we adopted in this work is shown in Algorithm~\ref{alg:Lanczos}.

\end{document}